\documentclass[runningheads]{cl2emult}

\usepackage{makeidx}  
\usepackage{graphicx} 
\usepackage{subeqnar} 
\usepackage{multicol} 
\usepackage{eso}      
\makeindex            

\begin{document}
\title*{Morphological Evolution of Galaxies to $z \sim 4$}
\toctitle{Morphological Evolution of Galaxies to $z \sim 4$}
%
%
\titlerunning{Morphological Evolution of Galaxies}
%
\author{Christopher J. Conselice\inst{1,2}}

\authorrunning{Christopher J. Conselice}
%
%

\institute{Space Telescope Science Institute, USA
\and University of Wisconsin-Madison, USA}

\maketitle              

\begin{abstract}
Galaxies have clearly evolved since the universe was 1 Gyr old, but methods 
to trace and quantify this evolution are still in their infancy.  In this
paper I demonstrate that with the careful use of a `physical morphology' it 
is possible to determine
quantitatively how the process of galaxy evolution is occurring out to $z
\sim 4$.  Using a system of parameters that traces star formation
and galaxy interactions, I show how distinct galaxy
populations at high-$z$ can be identified in deep high-resolution 
optical and NIR images.  These tools are also
used to measure a potential merger fraction of galaxies from 
$0<z<4$. If these methods are reliable, as is suggested by a local galaxy 
calibration, the merger fraction of galaxies scales 
$\alpha$ (1+z)$^{2.1\pm0.5}$, peaks near $z \sim 2$, and declines 
thereafter. I also discuss how this system is likely part of the
ultimate physical classification of galaxies.

\end{abstract}

\section{Introduction}

Although it is largely misunderstood and often misused, galaxy morphology 
is becoming a powerful tool for learning how galaxies 
form and evolve.   The {\it Hubble Space Telescope} now allows us to study
high-resolution `deep fields' of distant galaxies, yet
finding a way to utilize this information is not obvious or simple.

The dominate paradigm for understanding galaxies is still largely
based on the Hubble Sequence $[14]$ and its various 
revised forms.  Hubble classifications are useful
for describing the gross properties of nearby galaxies that are
for the most part in quiescent star formation and
dynamical states.  Even in this situation, Hubble types correlate
with physical properties, such as color and star formation rates, only
in the {\it mean} [21].  This is due to the fact that Hubble types
{\it do not} consider recent star formation in its classification, nor  
do interacting or merging properties factor into a Hubble type.  Because the 
merger rate of galaxies in the local universe is $<$ 3\% [20] and since
recent star formation for nearby galaxies is also low [18], the Hubble
sequence is effective in the mean for identifying nearby galaxy populations.

  The subjective Hubble morphological system 
is frankly inadequate at high redshift [e.g., 1,6].  Too many galaxies at 
high-$z$ fall in the catch-all ``irregular'' or ``peculiar'' class.   It is not
yet understood if these unusual looking galaxies are true irregulars forming
stars stochastically, or if they are starbursts triggered by some process,
perhaps galaxy mergers. In this paper, I will present preliminary work
showing how high-$z$ 
galaxies seen in deep optical and NIR images can be 
classified by understanding and quantifying star formation 
and merging properties based on their structures.  I also demonstrate how 
these tools can be used to perhaps trace the merger fraction of galaxies 
out to $z \sim 4$.

\section{Fundamental Parameters and Nearby Galaxy Calibration}

A truly physical morphology must correlate the way a galaxy `looks',
either through its structure or spectrum, with internal physical properties.
The physical morphological system being developed by Conselice et al. [5,6,7,8]
and Bershady et al. [2] thus far can distinguish the fundamental properties 
of star formation and galaxy interactions/mergers.  Figure 1a shows the 
color-asymmetry diagram (CAD), a diagnostic tool used to calibrate these 
properties using nearby galaxies including the Frei sample [12] 
supplemented by WIYN 3.5m
images of starburst galaxies.  The galaxies at $A>0.35$ are all consistent
with merging systems based on their broad HI line profiles [8].  The
starbursts with $A<0.35$, NGC 3310 and NGC 7678, have well defined
spiral patterns and whose starburst trigger was likely a bar instability,
or in the case of NGC 3310 a minor merger in the distant past [8].  The
CAD is therefore useful to quantify and trace the presence of not only
interacting galaxies, but ones undergoing mergers.
\vskip -0.75in
\begin{figure}
\centering
\hspace*{-0.55in}
\includegraphics[width=1.2\textwidth]{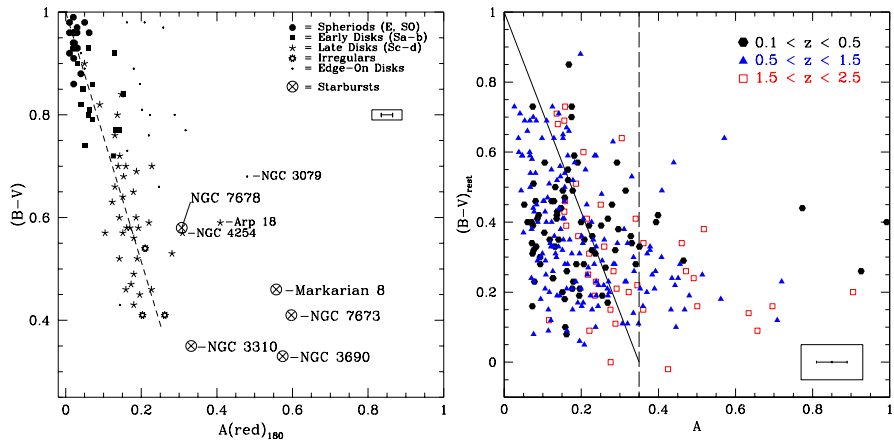}
\vskip -4in
\caption[]{a. Nearby galaxy color-asymmetry diagram from Conselice et al. [7,8]. Labeled galaxies are those consistent with an interaction or merger, with
all galaxies with A$>0.35$ having dynamical evidence for mergers [8]. b. 
Color-asymmetry diagram for galaxies in the HDF with A computed in the 
rest-frame B band.}

\label{eps1}
\end{figure}

The CAD is also useful for tracing star formation in galaxies, although
high-frequency structure filtering techniques will likely turn out to be more 
useful.  The CAD `fiducial' sequence, the dotted line in Figure 1a is likely 
a star formation sequence. As the star formation rate in a galaxy
increases, it becomes bluer and more asymmetric
due to the increase of localized star-forming regions. It is important to 
note that even galaxies whose
morphologies are dominated by star formation, such as the irregular NGC 4449,
do not have asymmetries that mimic the very high asymmetry produced
by merging [7,8].  Although this calibration is limited by the small 
samples used, the fact remains that the only galaxies with $A>0.35$
are mergers. This is consistent with other studies that 
demonstrate galaxies with the highest asymmetries are merging systems [23].

\section{The Evolution of Galaxy Populations to $z \sim 4$}

Do galaxies form by the merging of other galaxies?  How
important is rapid collapse in the galaxy formation process [19]?  The 
consensus
is that merging is the dominate process, as shown in both semi-analytic
and N-body modeling within Cold Dark Matter dominated cosmologies [15,22,4]. 
Finding a direct method of proving this
beyond circumstantial evidence, such as evolving luminosity functions at 
high-$z$ [16], remains a key observational challenge.

We can investigate this by the use of CADs measured at high-$z$.  Several
caveats must be stated before explaining the results of using quantitative 
features such as asymmetry at high-$z$.  An advantage for astronomers
studying galaxy evolution is the ability to study distant, and hence
young galaxies.  Thus we can directly compare properties of young and old 
galaxies to determine evolution.   This is easier said then done however, due 
to cosmological effects and biases.  When measuring the same parameters at 
high and low-$z$,
zeroth-order features such as magnitudes and colors can be measured with some
certainly by correcting for cosmological k-corrections and dimming.  Higher
order features, even simple measurements such as radii, are much harder
to measure reliably.  To robustly determine these features at high-$z$
{\em requires} simulating low-$z$ galaxies at high-$z$ to determine how
parameters change when images are degraded in resolution, and have
higher noise levels.
This is especially important for measuring morphological or
structural features. Conselice et al. [7] show that within certain limits, 
achievable with deep HST images, the asymmetry parameter measured on HDF 
galaxies with $m(B) < 25$ should reproduce the same values measured if the 
galaxies were nearby.

For the low-$z$ HDF range $0.1<z<0.5$ (Figure 1b), there are 
some galaxies with asymmetries consistent with mergers.  
There is also evidence for a possible correlation between asymmetry
and color as seen in the nearby galaxy sample.  There are no symmetric
red galaxies in this redshift range. This is largely the result of
a bias introduced by the small HDF field of view and the fact that 
the HDF was chosen to avoid any bright galaxies, that in the nearby
universe are typically red.  At the moderate redshift range $0.5<z<1.5$,
there are some red and symmetric galaxies (Figure 1b), and there is
an increase in the presence of systems consistent with merging.  At the
highest redshift range ($1.5<z<2.5$), we see an even higher fraction of 
interactions, and no galaxies consistent with red ellipticals.

\section{Merger Fractions}

A fundamental aspect for understanding galaxy evolution is to constrain
the history of galaxy merging in the universe.  Galaxy pair studies indicate 
that a high fraction of distant galaxies are likely interacting [e.g., 
20,17].  Dynamical pair fractions also reveal that the merger fraction of 
galaxies increases to at least $z = 1$
with a rate $\sim$ (1+z)$^{3.2}$ [17].  Dynamical pair studies are
limited by several factors, the least of which is the inability to
acquire spectroscopy for large numbers of faint galaxies.  Perhaps
a more important problem is the fact that these galaxies are not
necessarily merging systems, but might be grazing interactions or even galaxies
in orbits with several Gyr merger time scales.  Despite
this, dynamical pairs likely do measure some aspect of galaxy merging.

\begin{figure}
\centering
\vskip -0.3in

\includegraphics[width=.6\textwidth]{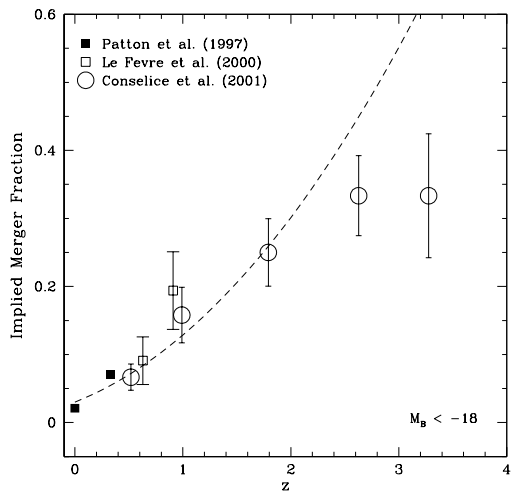}
\caption[]{Observed merger fraction evolution with redshift. The dashed line
shows the relationship f = 0.03 (1+z)$^{2.1}$, fitted out to z $\sim$ 2.}
\label{eps1}
\end{figure}
\vskip -0.2in

Using the asymmetry parameters on high-$z$ galaxies, it is possible to
demonstrate how {\em morphologically} the merger fraction at each
redshift evolves by
directly measuring the fraction of galaxies with structures
consistent with merging, as a function
of redshift. Doing this for galaxies in the HDF-N, we obtain Figure 2 which
shows the fraction of galaxies with rest-frame B-band asymmetries consistent 
with mergers as a function of $z$, using the criteria: $A>0.35$, 
M$_{B} < -18$. 
Also plotted on this figure is the merger fractions found from dynamical 
pair studies by Patton et al. [20] and Le F\'{e}vre et al. [17].  If we 
assume galaxies at these redshifts with $A>0.35$ are mergers, then the 
merger fraction increases as f = 0.03(1+z)$^{2.1\pm0.5}$ [10].
The merger fraction also potentially begins to decline at redshifts
higher than $z \sim 2$ [10].  

If the asymmetry parameter can indeed trace mergers out to these
redshifts, this preliminary observation is scientifically exciting, and 
suggests
that galaxy mergers are dominating the evolution of galaxies.  As is
well known, the star formation rate of galaxies peaks at about this
redshift [18,11], as does the presence of active galaxies [3].  Are these
effects and the mass assembly of galaxies driven by mergers?  This
morphological system suggests they are, demonstrating the power of
galaxy classification when used and calibrated properly.
  
\addcontentsline{toc}{section}{Acknowledgments}

This work would not have been possible without the active contribution
of many individuals, including the HDF-N NICMOS GO team, and especially
Mark Dickinson, Matt Bershady, Tamas Budavari, Jay Gallagher, and Anna
Jangren. 
I also appreciate support from a NASA Graduate Student Researchers Program 
(GSRP) Fellowship and from the STScI Graduate Student Program.

\clearpage
\addcontentsline{toc}{section}{Index}
\flushbottom
\printindex

\end{document}